\documentclass[12pt,letterpaper,natbib]{report}
\usepackage{amsmath,amsfonts,amssymb}
\usepackage[portuges]{babel} 
\usepackage[latin1]{inputenc}
\usepackage[T1]{fontenc}     	
\usepackage[dvips]{graphicx}
\usepackage{float}
\usepackage{epsfig}

\begin{document}

{\noindent \Large \bf Modelling spectral line profiles of wind-wind shock emissions from massive binary systems}
\medskip

{\bf D. Falceta-Gonçalves, Z. Abraham \& V. Jatenco-Pereira}

\medskip

{Instituto Astronômico e Geofísico - University of São Paulo}

\medskip
\begin{center}

\section*{Abstract}
\end{center}

One of the most intriguing spectral features of WR binary stars is the 
presence of time-dependent line profiles.  
Long term observations of several systems revealed the periodicity of this variability, synchronized 
with the orbital movement. 
Several partially successful models have been proposed to reproduce the observed 
data. The most promising assume that the origin of the emission is  the 
wind-wind interaction zone. In this scenario, two high velocity and dense winds produce a 
strong shock layer, responsible for most of the X-rays observed from  these systems.  As 
the gas cools down, flowing along the interaction surface, it reaches recombination 
temperatures and generates the emission lines. Luhrs (1997) noticed that, as the 
secondary star moves along its orbital path, the shock region, of conical shape, changes its position 
with relation to the line of sight. As a consequence, the stream measured Doppler 
shift presents time variations resulting in position changes of the spectral line.
However, his model requires a very thick contact layer and also fails to 
reproduce recently observed line profiles of several other WR binary systems. In our work, we
present an alternative model, introducing  turbulence in the shock layer to account 
for the line broadening and  opacity effects for the asymetry in the line profiles. 
We showed that the gas turbulence avoids the need of an unnaturally large 
contact layer thickness to reproduce line broadening. Also, we demonstrated that 
if the post-shock gas is optically thick at the observed line frequency, 
the emission from the opposing cone surface is absorbed, resulting in a 
single peaked profile. This result fully satisfy the recent data obtained 
from massive binary systems, and can help on the determination of both winds and 
orbital parameters. We succesfuly applied this model to the Br22 system and determined its orbital 
parameters. 
 
\vfill
      
\section{Introduction}

WR and O stars present mass-loss rates of $\dot{M} \simeq 5 \times 10^{-5} - 
10^{-7}$ M$_{\odot}$ yr$^{-1}$ and wind velocities of $v \simeq 2000 - 3000$ km 
s$^{-1}$ (Abbott et al. 1986, Willis 1991, Lamers \& Leitherer 1993, 
Lamers 2001). In  massive binary systems, the dense and supersonic winds produce a contact surface between 
two shock fronts along which the continuously incoming material flows.
The shocked gas is
characterized by  high densities and temperatures, the latter reaching  up to $T = 10^6 - 10^8$ K.

The shock region is detectable directly, by free-free continuum and line  emission 
from radio wavelengths to X-rays (Williams et al. 1990, Pittard et al. 1998, Bartzakos, 
Moffat \& Niemela 2001, Abraham et al. 2005), and indirectly by IR emission 
from dust formed at the contact surface (Monnier, Tuthill \& Danchi 2001, Falceta-Gon\c calves, 
Jatenco-Pereira \& Abraham 2005). 

Isolated hot stars present  broadened emission lines 
with variability in both amplitude and profile, which are usually  attributed to the clumpiness of the 
stellar winds 
(L\`epine \& Moffat 1999, L\`epine et al. 2000). However, in binary systems, the origin of the variability 
must be 
related to the binary system properties, since the observed line 
profiles repeat each orbital cycle (Seggewiss 1974). Luhrs (1997) proposed 
a purely geometrical model in which the observed spectrum  was a composition 
of the stellar and the shocked gas emissions. This model was 
quite successfully applied to WR 79, showing itself very useful for the determination of 
the orbital parameters, as period and eccentricity, as well as  stellar 
wind velocities and mass loss rates. However, 
 Hill et al. (2000) and Bartzakos, Moffat \& Niemela (2001) 
showed that the model was not able to fully reproduce the line profiles of several other WR binary 
systems.  

In this work, we present a similar but improved model, in which we also assume  that the lines are 
produced in the post-shock gas,  but their profiles are affected by turbulence  and opacity  in the  
contact surface, which change  at each point of the binary orbit. In section 2 we describe the 
model geometry and flux calculation, in section 3 
we present the  
results obtained for different stellar and orbital parameters, followed by the 
conclusions in section 4.

\section{The Line Profile Model}

To model the line profile it is necessary to determine the relationship  
between the line intensity and the observed velocity. Since we assume that the line is produced in the 
shocked region, it is necessary to know the temperature structure and  gas flow velocity  along the 
contact surface projected into the line of sight for each orbital phase. We will present next a simplified 
model for the shock geometry and discuss latter the calculation method for the line profile including 
turbulence and opacity.

\subsection{The geometry}

When the two winds collide, an interaction zone is created that consists 
of two shock fronts at both sides of a contact surface. At the shock 
fronts, the temperature of the gas is highly increased, reaching  up to 
$10^5 - 10^8$ K, depending on the wind velocities. The densities are also 
higher in the interaction zone. For adiabatic supersonic shocks, the density 
increases by a factor $\sim 4$, but if radiative cooling is efficient, this 
factor may be much larger.

The contact surface is defined as the region where the momentum balance 
between the two winds occurs. Luo, McCray \& Mac-Low (1990) and Stevens, Blondin \& Pollock (1992) presented an analytical equation for 
its geometry, given by:

\begin{equation}
\medskip
\frac{dy}{dz}= \frac {(\eta^{-1/2}{d_2}^2+ {d_1}^2)y}{\eta^{-1/2}{d_2}^2z+{d_1}^2(z-D)},
\medskip
\end{equation}

\noindent
where $D$ is the distance between the  stars; $d_1$ and $d_2$ are 
the distances of the primary and secondary stars to the 
contact surface, respectively, and $\eta=\dot{M_s} v_s/\dot{M_p} v_p$, 
where  $\dot{M_p}$ and $\dot{M_s}$ are the mass
 loss rates of primary and the secondary stars, and
$v_p$ and $v_s$ their respective wind velocities. 

Asymptotically, this 
surface becomes conical, with an opening angle $\beta$. In Figure 1 (left 
panel) we show the contact surface  ($\rm S_C$) given by Equation 1, for $\eta = 0.1$ 
that corresponds to $\beta \simeq 44^\circ$;  the two shock fronts $\rm S_1$ and $\rm S_2$ are also
schematically represented. For certain recombination lines (e.g. UV and visible) we expect the major 
emission to come from larger distances of the shock apex, where the gas has cooled. In 
this case, 
we will assume the conical geometry in the further calculations.

\begin{figure*}
\centering
\includegraphics[width=14cm]{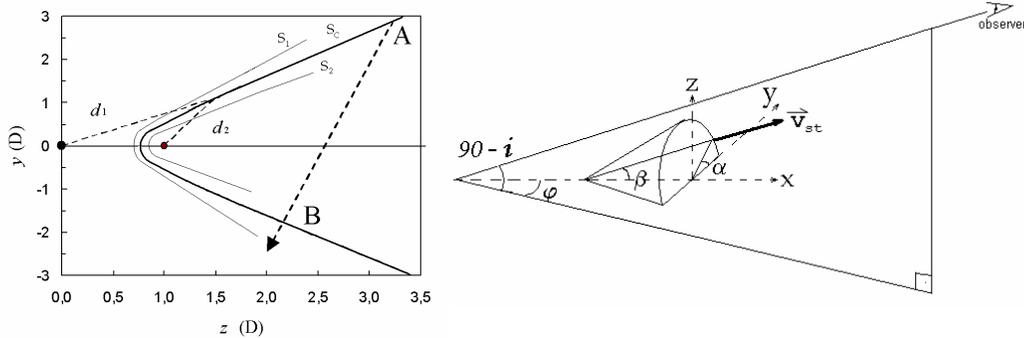}
\caption {Left: Momentum balance surfaces of the colliding winds for 
both $\eta = 0.1$ which result in $\beta = 44^\circ$. The dots represent each star, and the 
arrow represents the line of sight intercepting the points labeled A and B ($see$ Section 2.3). Right: 
Schematic view of the assumed geometry.}
\label{figure1}
\end{figure*}
 
We will assume that all the physical parameters of the shock gas 
are cylindrically  symmetric with respect to the cone axis and that 
the stream velocity is constant along the contact surface. As we 
will see later, with these assumptions it is not necessary to know 
the exact temperature and density distributions of the shocked material to calculate the 
normalized line profiles. 

To calculate the  stream 
velocity projected into the line of sight we use the schematic geometry shown 
in the right panel of Figure 1. We define the $x$ axis by 
the straight line containing both stars and, as a consequence, coincident to the cone
symmetry axis. Notice that this is correct if inertial deformations due to 
the orbital motion of the secondary star are neglected. 
We will discuss latter the main effects of this deformation in the 
obtained results. The $y$ axis, together with $x$, defines the orbital plane.  
  
The  stream velocity projected over the line of 
sight, as shown in Figure 1, is given by:

\begin{eqnarray}
\medskip
v_{\rm obs} = v_{\rm flow} (-\cos \beta \cos \varphi \sin i + \sin \beta \cos 
\alpha \sin \varphi \sin i \nonumber \\
- \sin \beta \sin \alpha \cos i),
\medskip
\end{eqnarray} 

\noindent
where $v_{\rm flow}$ is the stream velocity, $i$ is the inclination of the orbital plane  with respect to 
the line 
of sight, $\alpha$ is the cone azimuthal angle and $\varphi$ is the orbital phase, 
defined here as the angle between $x$ and the projection of the 
line of sight on the orbital plane.

Equation 2 shows that, for a given orbital phase 
$\varphi$, each fluid element with a different $\alpha$ value will contribute with a different Doppler 
shift.

 Assuming azimuthal symmetry for the gas properties over the cone surface, the line intensity distribution 
$I(\alpha)$ originated by a volume element at azimuth $\alpha$ will be:

\begin{equation}
\medskip
I(\alpha) = \frac{I}{2 \pi}, 
\medskip
\end{equation}

\noindent
where $I$ is the 
total line intensity. The line profile distribution $I(v_{\rm obs})$ 
can be obtained from the relation:

\begin{equation}
\medskip
I(\alpha) d\alpha = I(v_{\rm obs}) dv_{\rm obs}. 
\medskip
\end{equation}
 
Substituting $dv_{\rm obs}/{d\alpha}$, given by Equation 2 into Equation 4, the line 
profile can be  determined. In Figure 2 we show the calculated synthetic line profiles  for 
different values of the orbital phase, assuming $i = 70^\circ $ and $\beta = 40^\circ $. 

\begin{figure}
\centering
\includegraphics[width=7cm]{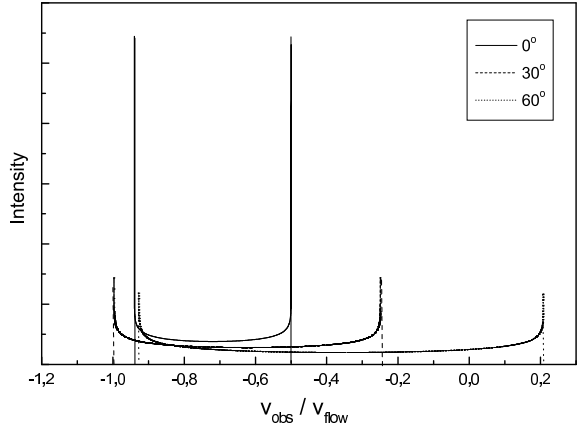}
\caption {Synthetic line profiles, obtained from Equations 2 - 4 for 
orbital phases ranging from $0^\circ$ to $180^\circ$, assuming $i = 70^\circ$ and $\beta = 40^\circ$.}
\label{figure2}
\end{figure}

\subsection{Line Broadening}

The profiles shown in Figure 2 are unable to 
reproduce the observations because they are discontinuous and present  large differences
between the peak and the lower emission intensities. The introduction of some sort of 
broadening could reduce the discrepancy.

Luhrs (1997) noticed from numerical simulations of adiabatic shocks that the emitting region is delimited 
by two conic  surfaces with $\beta_1 < \beta < \beta_2$.  
Using an $ad$ $hoc$ flux density distribution over $\beta$ he obtained  broadened synthetic line profiles. 
Varying  $\Delta \beta = \beta_2-\beta_1$ between 37$^\circ$ and 
54$^\circ$, depending on the orbital phase, he was able to  fit for WR79 the observed CIII 5696\AA \ lines 
at some orbital phases,  but failed to reproduce others. Besides, in massive 
binary systems,  the shock 
is highly radiative because of the high density at the contact surface, resulting in thin and turbulent 
layers (Stevens, 
Blondin \& Pollock 1992). For that reason,  $\Delta \beta$ is  expected to be lower 
near pariastron, as the density is higher, in disagreement to the hypothesis used by Luhrs (1997). The 
variable $\Delta \beta$ model  also failed to reproduce  the line profiles 
 of other WR binary systems (Hill, Moffat, 
St-Louis \& Bartzakos 2000, Bartzakos, Moffat \& Niemela 2001).

Since the numerical models mentioned above  predict the existence of a highly turbulent contact surface, 
in the present work we introduced turbulence in the calculation of the line profile, using a constant 
single value for the $\beta$ parameter. For that, 
we assumed that each emitting element at a given $\alpha$ has a gaussian velocity distribution, 
with a mean value obtained from Equation 2, and a dispersion amplitude 
$\sigma$. Therefore, different emitting elements, at different $\alpha$ values 
may present the same observed velocity. In this case, since there is no univocal 
relationship between the velocity projected into the line of sight and 
$\alpha$, Equation 4 is no longer valid and the line profile must be obtained from:

\begin{equation}
\medskip
I(v) = \mathcal{C} \int_{0}^{\pi }\exp \left[ -\frac{\left( v-v_{\rm obs}\right) ^{2}}{2\sigma
^{2}}\right] d\alpha, 
\medskip
\end{equation} 

\noindent
where $\mathcal{C}$ is the normalization constant. Substituting the mean value $v_{\rm obs}$, given by 
Equation 2, and 
integrating the right hand side of the equation over $\alpha$, we obtain 
the relative line intensity for each projected velocity into the line of sight $v$.

\subsection{Opacity Effects} 

Bartzakos, Moffat and Niemela (2001) 
obtained the spectra of Br22 (WC4), a binary system in
the Magellanic Clouds, at several epochs. They found that the CIII 5969\AA \ line could be explained by a 
superposition of a constant feature  and an emission excess that varied across the line profile according 
to the orbital phase. In 
contrast to what is expected for binary systems from the Luhrs model, 
these excesses were single peaked. Norci et al. (2002)  observed 
the same behavior for several other Galactic WR stars.

A possible explanation for the single peaked profiles 
could be the enhanced absorption of some of the emission 
by the dense gas layer at the contact surface, when positioned in front of it.
In fact, the density of the interaction zone  depends on the 
stellar mass-loss rates and on the  separation between the stars, and 
could eventually become high enough to absorb part of the incoming photons (Falceta-Gon\c calves, 
Jatenco-Pereira \& Abraham 2005).    

In Figure 1 (left pannel), we illustrate a given situation in which the 
line of sight (dashed arrow) intercepts the regions of the contact surface labeled 
A and B. As the cone is intercepted 
by the line of sight, the emission element A, responsible for the 
``redder" peak of the synthetic profile, could be obscured by the absorbing element B.
The opacity of this element  is unknown, since it 
depends both on the undetermined gas density and temperature. 
However, if the cooling process 
is effective, the gas will 
reach high densities as the temperature diminishes, becoming eventually optically 
thick for certain wavelengths.  This is expected to occur mainly in close 
massive binary systems. 

In this work we introduced an optical depth $\tau$ into
Equation 5. For a given orbital phase, when two emitting 
fluid elements  are intercepted by the line of sight, 
the observed emission from the farthest element will be diminished by a factor $\exp(-\tau)$ while the 
emission from the nearest will remain unchanged. If $\tau >> 1$, we could obtain
a single peaked line profile. Actually, a more detailed model solving the complete radiative 
transfer equations would result in a lower relative intensity for the intervening material emission. 
However, the qualitative result for $\tau >> 1$ remains the same, and for $\tau < 1$ no 
important changes occurs. 

Since the opacity of the contact surface  depends on the separation between the stars, 
the interaction region of close binary systems could be optically thick 
while systems with large separations  could be optically thin. That explains why certain systems present a 
single peak in their 
spectra during all the orbital period, while others present double peaked 
profiles.

\section{Results and Discussion}

We will investigate the effects of the binary system parameters on the line profiles, using the model
described in section 2. Initially, we will neglect 
opacity, $i.e.$ use $\tau = 0$, fix 
$\beta = 40^\circ$ and a low turbulence amplitude $\sigma / v_{\rm flow} = 0.05$,  
varying the orbital inclination $i$. In Figure 3 we present the normalized line 
intensity profiles,  in arbitrary units, for 
$i = 90^\circ$ (top), $70^\circ$ (middle) and $20^\circ$ (bottom), at different orbital 
phases from $\varphi = 0^\circ$ to 180$^\circ$. For $i = 90^\circ$, at $\varphi = 0^\circ$, 
the cone is pointed directly towards the observer with all emitting gas 
being observed with the same mean velocity, resulting in a strong 
peak with mean velocity given by the first term of Equation 2.  
The highest Doppler shift occurs when $\varphi = \beta$ and 
$\varphi = 180^\circ - \beta$, as the line of sight coincides with the 
cone generatrix. Also, it is noticeable that the 
line profiles present larger variation during the orbital period for larger values of $i$. For face-on
oriented systems, the observed line will not show any variation during 
the orbital period.

\begin{figure}
\centering
\includegraphics[width=8cm]{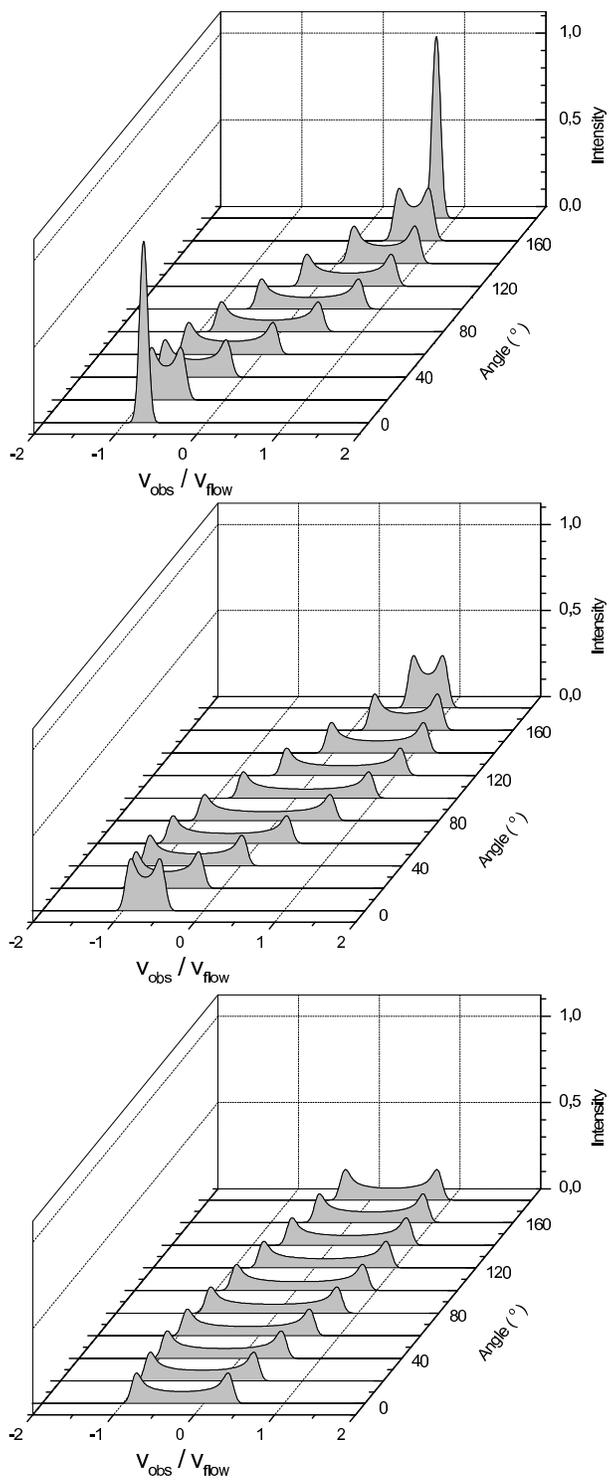}
\caption{Theoretical line profiles considering $\sigma / v_{\rm flow} = 0.05$ and 
$\beta = 40^\circ$ for $i = 90^\circ$ (top), $70^\circ$ (middle) and $20^\circ$ (bottom), 
neglecting absorption.}
\label{figure 3}
\end{figure}

In Figure 4 we show the effects of the cone opening angle on the line 
profiles. Here we neglected absorption, fixed $i = 70^\circ$ and 
$\sigma / v_{\rm flow} = 0.05$, and compared the results for 
$\beta = 40^\circ$ (top) and $70^\circ$ (bottom) at different orbital phases. The main difference 
 occurs on the distance between the peaks. The larger 
the opening angle the larger is the peak separation. It shows that, 
for a given system in which both peaks are detectable during all orbital 
phases, it is possible to determine $\beta$, $i.e.$ the relation between 
the stellar wind momenta.

\begin{figure}
\centering
\includegraphics[width=8cm]{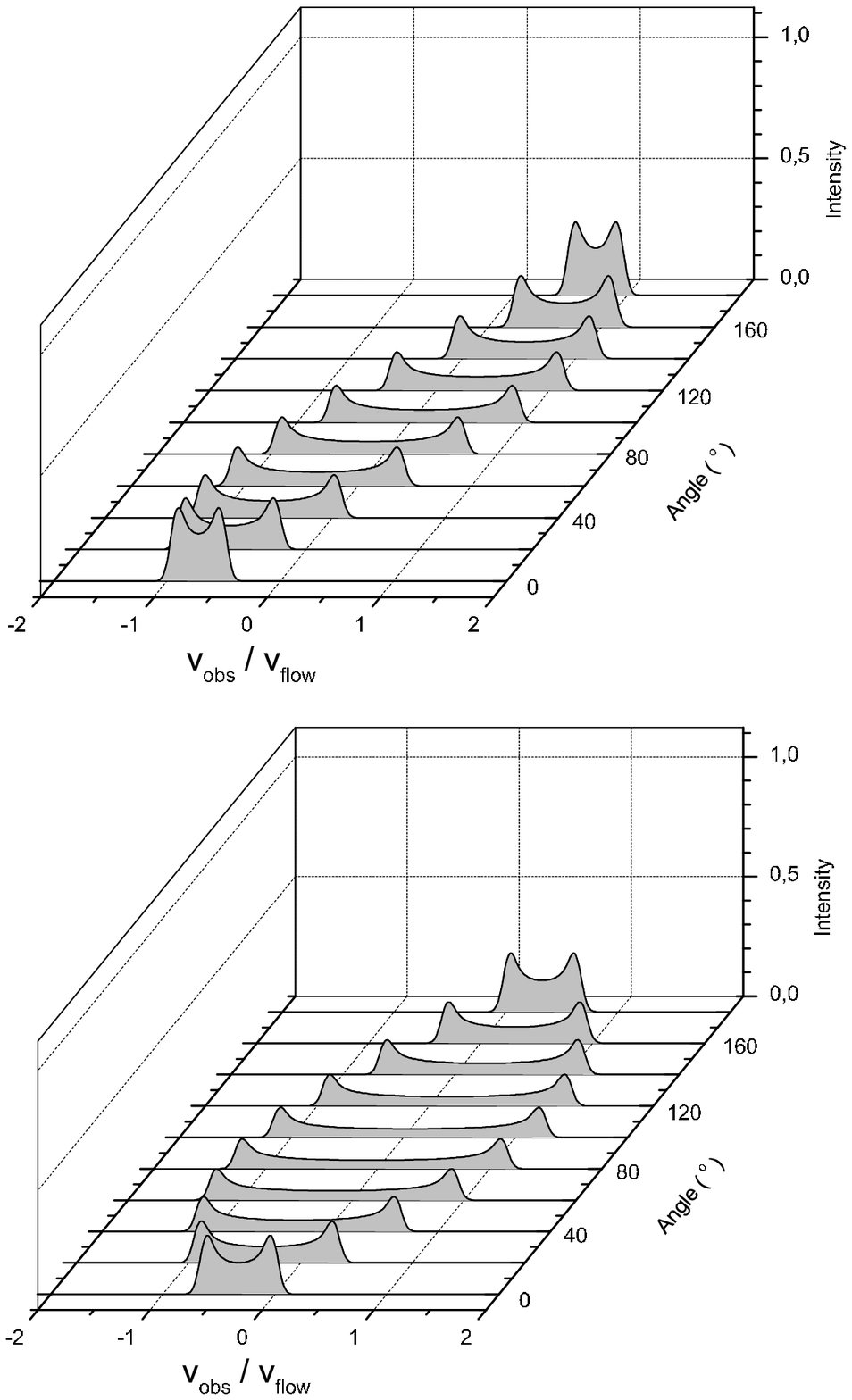}
\caption 
{Theoretical line profiles considering $i = 70^\circ$ and 
$\sigma / v_{\rm flow} = 0.05$ for $\beta = 40^\circ$ (top) and $70^\circ$ (bottom), 
neglecting absorption.}
\label{figure4}
\end{figure}

The stream turbulence also play an important role in the emission line 
profiles. In Figure 5, we neglected absorption, fixed $i = 70^\circ$ and 
$\beta = 40^\circ$, and obtained the results for the turbulence amplitudes 
$\sigma / v_{\rm flow} = 0.05$ (top) and 0.20 (bottom) at different orbital 
phases. Obviously, by increasing 
the flow turbulence at the post-shock region we obtained broadened line 
profiles, as in Luhrs (1997). However, in this case, it was not necessary 
to use a high and unrealistic difference between the internal and external 
cone opening angles. A high turbulent system may, eventually, show a 
plateau instead of a two peaked profile as both peaks tend to merge. This would 
be an indication of a strongly radiative shock.

\begin{figure}
\centering
\includegraphics[width=8cm]{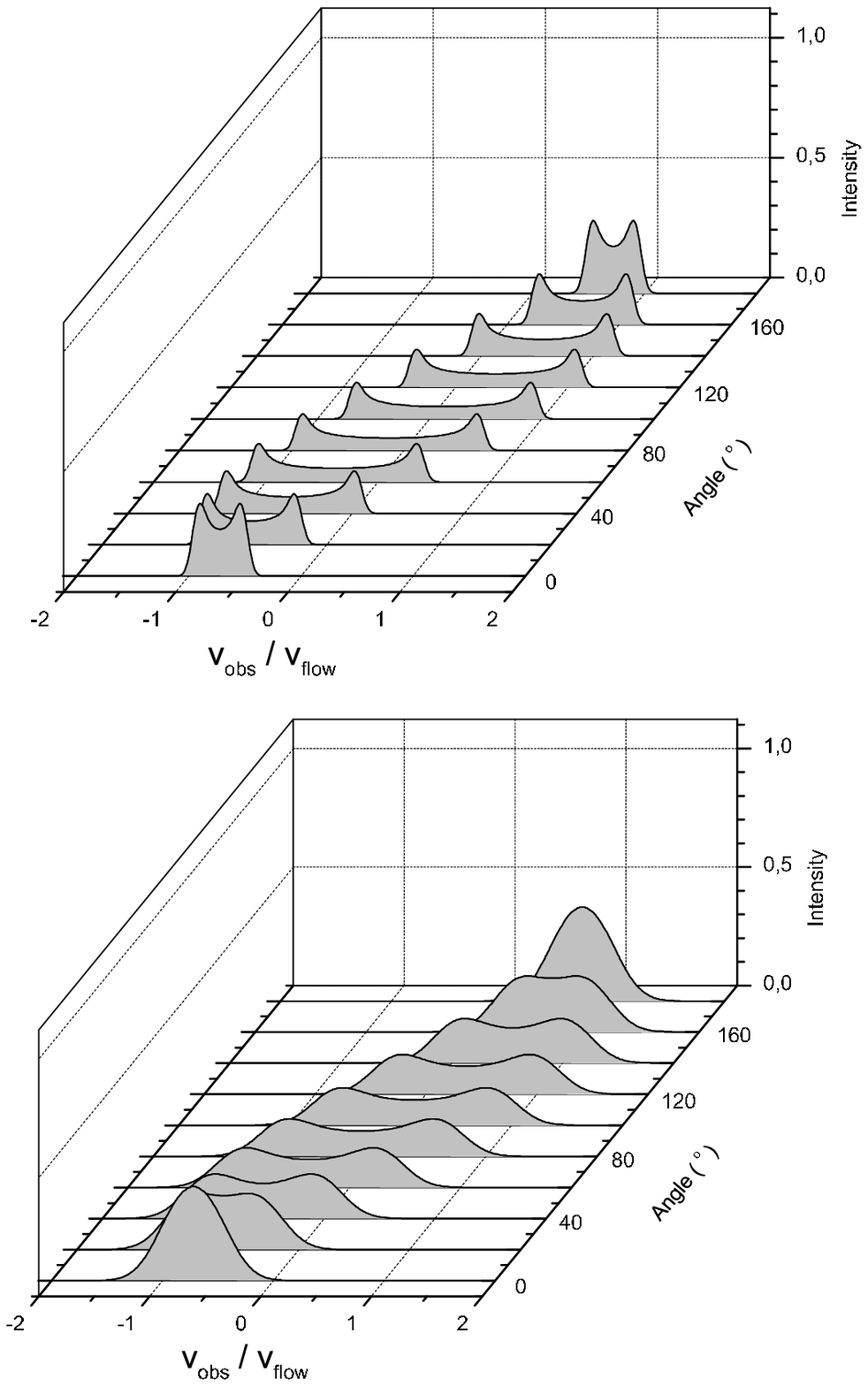}
\caption {Theoretical line profiles considering $i = 70^\circ$ and 
$\beta = 40^\circ$ for $\sigma / v_{\rm flow} = 0.05$ (top) and $0.20$ (bottom), 
neglecting absorption.}
\label{figure5}
\end{figure}

\begin{figure}
\centering
\includegraphics[width=8cm]{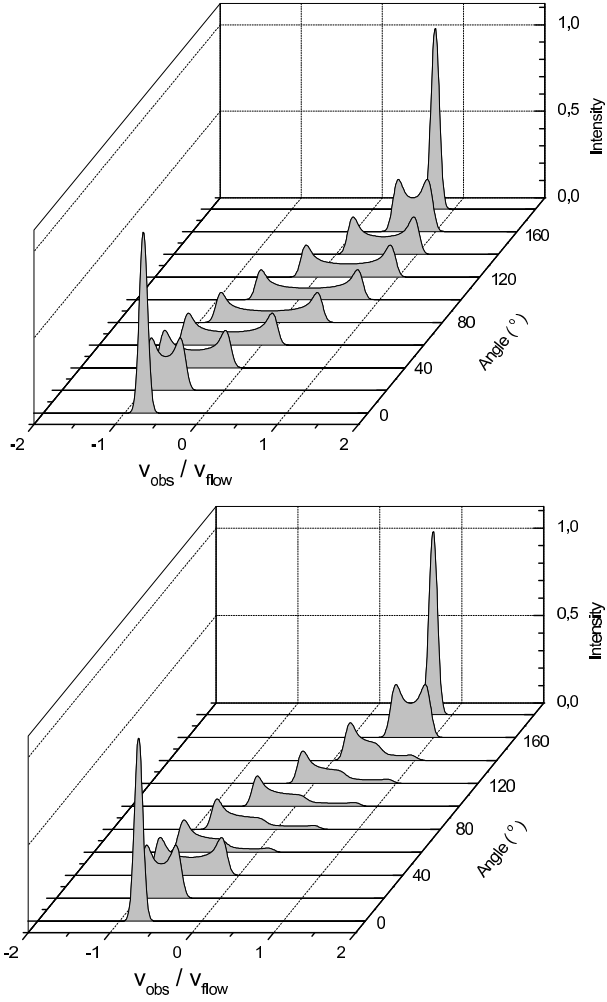}
\caption {Theoretical line profiles considering $\sigma / v_{\rm flow} = 0.05$, 
$\beta = 40^\circ$ and $i = 90^\circ$ for $\tau = 0$ (top) and $\tau >> 1$ (bottom).}        
\label{figure6}
\end{figure}

Finally, in Figure 6 we simulate a strong absorption during the synthetic 
line profile computation. For that, we fixed $i = 90^\circ$, $\beta = 40^\circ$ and 
$\sigma / v_{\rm flow} = 0.05$, and performed the calculations for $\tau = 0$ (top) and 
$\tau >> 1$ (bottom) at different orbital 
phases. In this case,  the main difference  in the line profile
occurs at the orbital phases $\beta < \varphi < 180^\circ - \beta$, in which the redder part of the 
profile is 
absorbed, resulting in a single peaked profile. 
For face-on systems ($i = 0^\circ$), the single peak feature occurs during 
the entire orbital period. Also, an interesting fact is that the 
blueshifted peak will be unabsorbed, while the redshifted may be totally 
absent. Qualitatively, this result is in full agreement with the observations 
of several binary systems that were not reproduced by the Luhrs synthetic 
profiles (Bartzakos, Moffat and Niemela 2001, Norci et al. 2002).   

\begin{figure}
\centering
\includegraphics[width=6cm]{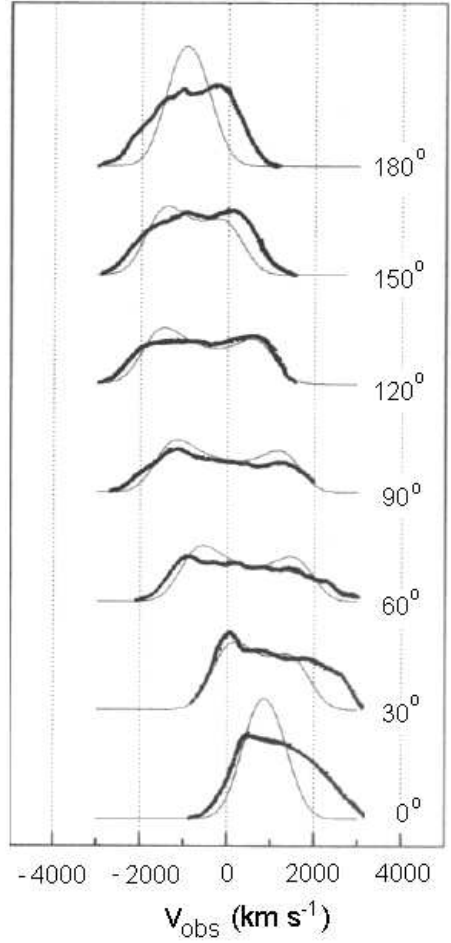}
\caption {Model application for Br22 assuming $i = 71^\circ$, 
$v_{flow} = 1500$ km s$^{-1}$ and $\beta = 47^\circ$ 
(Bartzakos, Moffat \& Niemela 2001), and adjusting 
$\sigma / v_{flow} = 0.25$ and $\tau = 0.6$.}        
\label{figure7}
\end{figure}

\subsection{The case of Br22}

Br22 (HD 35517), a massive WR+O binary system located in the Large Magellanic Cloud, presents  periodic 
line profile variations. The CIII 5696 \AA \ 
emission line is probably produced at the shock site and can provide clues 
on  the wind and orbital parameters. 
Bartzakos, Moffat \& Niemela 
(2001) applied Luhrs model to this object, obtaining a best fit 
for $i = 30^\circ$, $v_{flow} 
= 2700$ km s$^{-1}$ and $\Delta \beta = 45^\circ$. However, photometric, 
spectrometric and polarimetric studies determined the actual orbital inclination as 
$i \sim 70^\circ$; and peak position analysis gave  a direct estimate for the 
stream mean velocity $v_{flow} \sim 1500$ km s$^{-1}$. Also, qualitatively, the 
observed peak relative intensities are in disagreement to the synthetic lines. 
These discrepancies show that the Luhrs profiles are not able to reproduce these observations.

We applied the model described in this work for the Br22 
system fixing $i = 71^\circ$, $v_{flow}, \sim 1500$ km s$^{-1}$ and 
$\beta = 47^\circ$, as determined by  the observations. Then, we varied the turbulence 
amplitude and the opacity in order to obtain the best fitting to the observed profiles (Figure 5 of 
Bartzakos, Moffat \& Niemela 2001). This result, corresponding to  $\sigma / v_{flow} = 0.25$ and 
$\tau = 0.6$, is shown in Figure 7, where the light and dark lines correspond to the synthetic 
and the observed profiles, respectively. As explained above, since the two peaks are marginally detected 
from the observations, the value of $\tau$ must be lower than unity, as determined by the 
best fitting.

As we can see, the synthetic profiles match quite 
well both the bulk profile and the peak positions. Although considerably better than the 
Luhrs model adjusts, the bottom spectra of Figure 7 show broader observed lines. Actually, 
there is some diffulties on separating the emission lines from the continuum, and
better fittings could be obtained depending on the data baseline subtraction. If real, these 
differences can be explained if, instead of a perfect cone, the shock surface is curved 
by the stellar orbital motion. In this scenario, the emission would not be concentrated to 
form the two peaked profile, but distributed over the possible radial 
velocities. Also, we assumed that the flow velocity is constant during the orbital period what 
is not necessarily true. Higher eccentric orbits would lead to different flow velocities, and 
different shock turbulence, as the stars move along their orbits. 
These hypothesis could be verified using time-dependent numerical simulations.

In addition to the turbulence and opacity derived from our model, we were able to 
constrain some of the orbital parameters. To do that, for each orbital phase 
obtained from the observations, we determined the corresponding model phase angle, 
which depends on the eccentricity ($e$) and the angle between the semi-major axis and 
the line of sight ($\varphi_0$), by comparing the shape of the line profile and the peak position. The 
results of our best fitting are  $e = 0.1$ and $\varphi_0 = 35^\circ$; in Figure 8 
we show the dependence of the phase angle on the orbital phase for three 
values for the eccentricity (left panel) and the best fit to the observation 
(right panel).

\begin{figure*}
\centering
\includegraphics[width=14cm]{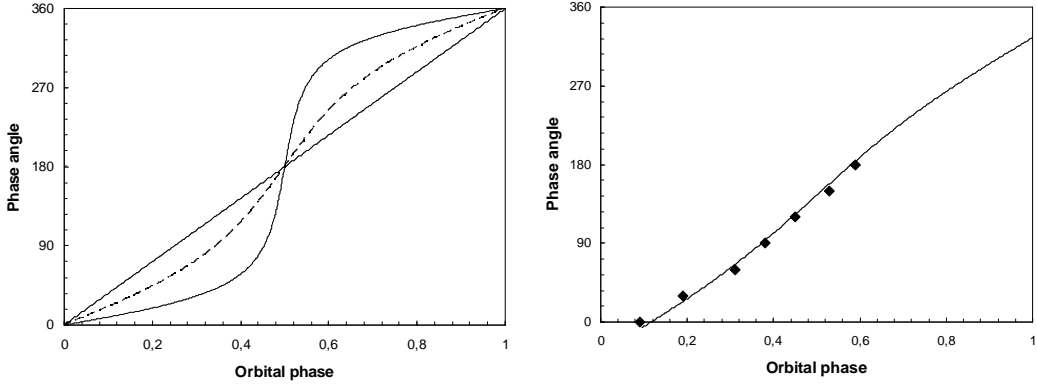}
\caption {Left: The dependence of the phase angle on the orbital phase for
eccentricities e = 0.0 (solid), 0.3 (dashed) and 0.7 (dotted). Right: the
phase angle variation, obtained from Figure 7, for Br22 and the best fit
for $\varphi_0 = 35^\circ$ and e = 0.1.}        
\label{figure8}
\end{figure*}

\section{Conclusions}

In this work we present a model to explain the variable
line emission profiles of WR binary stars. Following Luhrs (1997), we 
assumed that the observed emission lines  are formed  at the wind-wind interaction zone by the cooling 
gas flowing along the contact surface. Besides, we introduced 
turbulence as a new parameter and also studied the effects of the shock layer 
opacity on the synthetic line profiles.

Our model recovered the standard result of  two peaked profiles, and their
 dependence  on the orbital plane inclination and on the cone opening 
angle. However, the line broadening is assumed to be the consequence of a 
highly turbulent environment in strongly cooling shocks. In this case, 
it is not necessary to assume unreal large shock layer widths to reproduce 
the observations. We also showed that the  opacity is responsible for the intensity reduction 
of the  reddened peak. In this sense, a highly opaque post-shock gas results 
in a line profile with a single blue-shifted peak, as occurs for some 
observed systems.

In particular we applied this model to Br22 and were able to reproduce fairly well the line profiles for 
the different orbital phases using the observed parameters $i = 71^\circ$, $v_{flow}, \sim 1500$ km 
s$^{-1}$ and $\beta = 47^\circ$ and assuming $\sigma / v_{flow} = 0.25$ and 
$\tau = 0.6$. By comparing the observed and model profiles we were also able to determine the eccentricity 
and the angle between the periastron and the line of sight ($e = 0.1$ and $\varphi_0 = 35^\circ$).

We did not take into account the inertial effect of the secondary motion 
on the cone geometry deformation. A direct effect is the variation on 
the angle between the cone axis and the line of sight. However, we performed 
our calculations in terms of $\varphi$, which exactly corresponds to this 
parameter and not strictly the orbital phase. Actually, to obtain the orbital 
phase it is necessary to subtract (or add) the angle variation 
($\delta \varphi$) to $\varphi$. The other effect of the stellar motion is 
the curvature of the cone generatrix. As discussed above, in the case of very
close binaries, this curvature could be high enough to flatten the observed 
profile. We also did not take into account the possible phase-dependent variations of the 
flow velocity and shock turbulence, as could occur in high eccentric orbits.
Numerical simulations are needed to quantify these effects. 

\section*{Acknowledgments}

D.F.G thanks FAPESP (No. 04/12053-2) for financial support. Z.A. and V.J.P. thank 
FAPESP, CNPq and FINEP for support. 
  
\section*{References}
\noindent
Abbott, D. C., 
Bieging, J. H., Churchwell, E. \& Torres, A. V. 1986, ApJ, 303, 239. 

\noindent
Abraham, Z., 
Falceta-Gon\c calves, D., Dominici, T. et al. 2005, MNRAS, 364, 922.

\noindent
Bartzakos, P., 
Moffat, A. F. J. \& Niemela, V. S. 2001, MNRAS, 324, 33.

\noindent
Davidson, K.,  
Martin, J. C., Humphreys, R. M. \& Ishibashi, K. 2005, AAS, 206, 808.

\noindent
Falceta-Gon\c calves, 
D., Jatenco-Pereira, V. \& Abraham, Z. 2005, MNRAS, 357, 895.

\noindent
Henley, 
D. B., Stevens, I. R. \& Pittard, J. M. 2003, MNRAS, 346, 773.

\noindent
Hill, G. M., 
Moffat, A. F. J., St-Louis, N. \& Bartzakos, P. 2000, MNRAS, 318, 402.

\noindent
Hummer, D. G. \&
 Storey, P. J. 1987, MNRAS, 224, 801.

\noindent
Lamers, H. J. G. L. M.
\& Leitherer, C. 1993, ApJ, 412, 771.

\noindent
Lamers, H. J. G. L. M. 2001, PASP, 
113, 263.

\noindent
L\`epine, S. \& 
Moffat, A. F. J. 1999, ApJ, 514, 909.

\noindent
L\`epine, S., 
Moffat, A. F. J., St. Louis, N., Marchenko, S. V. et al. 2000, AJ, 120, 3201.

\noindent
Luhrs, S. 1997, 
PASP, 109, 504.

\noindent
Luo, D., McCray, R. \& Mac-Low, M. 1990, ApJ, 362, 267.

\noindent
Monnier, J., 
Tuthill, P. \& Danchi, W. 2001, A\&AS, 199, 608.

\noindent
Norci, L., 
Viotti, R. F., Polcaro, V. F. \& Rossi, C. 2002, RMxAA, 38, 83.

\noindent
Pittard, J. M., Stevens, 
I. R., Corcoran, M. F. \& Ishibashi, K. 1998, MNRAS, 299, 5.

\noindent
Seggewiss, W. 1974, 
A\&A, 31, 211.

\noindent
Stevens, I. R., 
Blondin, J. M. \& Pollock, A. M. T. 1992, ApJ, 386, 265.

\noindent
Williams, P. M., 
van der Hucht, K. A., Pollock, A. M. T. et al. 1990, MNRAS, 243, 662.

\noindent
Willis, A. J. 1991, Proc. IAU 143, Wolf-Rayet 
Stars: and 
interrelations with other massive stars in galaxies, ed. K. A. van der Hucht 
and B. Hidayat (Dordrecht, Kluwer), p265.
 

\end{document}